\documentclass[twocolumn]{aastex62}
\usepackage{xfrac}

\submitjournal{ApJ}

\shorttitle{Hierarchical PM}
\shortauthors{Gnedin}

\def\vec#1{\mathbf{#1}}
\def\dvec#1{{\rm d}\vec{#1}}
\def\half{\sfrac{1}{2}}

\def\oct#1{\sfrac{#1}{8}}

\begin{document}
\title{Hierarchical Particle-Mesh: an FFT-accelerated Fast Multipole Method}

\correspondingauthor{Nickolay Y.\ Gnedin}
\email{gnedin@fnal.gov}

\author{Nickolay Y.\ Gnedin}
\affiliation{Fermi National Accelerator Laboratory;
Batavia, IL 60510, USA}
\affiliation{Kavli Institute for Cosmological Physics;
The University of Chicago;
Chicago, IL 60637 USA}
\affiliation{Department of Astronomy \& Astrophysics; 
The University of Chicago; 
Chicago, IL 60637 USA}

\begin{abstract}
I describe a modification to the original Fast Multipole Method (FMM) of Greengard \& Rokhlin that approximates the gravitation field of an FMM cell as a small uniform grid (a "gridlet") of effective masses. The effective masses on a gridlet are set from the requirement that the multipole moments of the FMM cells are reproduced exactly, hence preserving the accuracy of the gravitational field representation. The calculation of the gravitational field from a multipole expansion can then be computed for \emph{all multipole orders} simultaneously, with a \emph{single} Fast Fourier Transform, significantly reducing the computational cost at a given value of the required accuracy. The described approach belongs to the class of "kernel independent" variants of the FMM method and works with any Green function.
\end{abstract}


\section{Introduction} \label{sec:intro}

The Fast Multipole Method (FMM) invented by Greengard \& Rokhlin \citep{gr87,gr97} is deservedly quoted as one of the top 10 computational algorithms of the XX century \citep{cipra00}. It is the only known method for computing self-gravity of a collection of bodies ("particles") that has asymptotically linear scaling with the number of interacting particles. It does, however, have a bottleneck - the key step of the algorithm, the computation of the gravitational field from the multipole expansion is rather expensive. Not surprisingly, significant effort has been put into accelerating this step of the algorithm.

The FMM algorithm relies on the subdivision of the whole computational domain into separate cells organized in an oct-tree in 3D or a quad-tree in 2D. The FMM algorithm and its variant described here are not dimension specific, however, for the sake of presentation, whenever a specific value for the dimension of space is needed, I will choose the 3D case as the most common case for astrophysical applications. While there is significant freedom in specifying the details of the underlying tree, in this paper I only consider a cubic (square in 2D) computational domain and cubic oct tree cells. With this restriction, the whole computational domain forms the single cell on the level 0 (or root level). The level 0 cell is divided in 3D into 8 "child" cells (into 4 cells in 2D and 2 cells in 1D), and becomes their "parent" cell. Each of the child cells can, in turn, be divided into 8 more child cells, etc. The resultant tree may or may not be of uniform depth, hence supporting Adaptive Mesh Refinement.

The algorithm itself consists of 4 main steps. In the first step the gravitational field (expressed either as gravitational potential or accelerations due to gravity) of each childless cell - a "leaf" - is computed and parameterized with a finite number of parameters. In the original FMM formulation of Greengard \& Rokhlin \citep{gr87,gr97} these parameters are the multipoles of the matter distribution, but other forms of parameterizations have been used as well and it is also the key ingredient of the Hierarchical Particle-Mesh  method described in this paper. In the second step the parameterizations of the child cells are combined into a parameterization of their parent cell for all cells in the tree. In the original FMM formulation this step was called "multipole-to-multipole" or M2M translation. To be more general and allow for non-multipole parameterizations, I am going to call it the "child-to-parent transformation". At the end of the child-to-parent transformation, each cell in the tree has a parameterized approximation to the gravitational field created by all the matter inside it.

The third step is the key step of the algorithm, and relies on the subdivision of space around a given cell in the tree (I will call it a "target" cell) into two zones: the "near zone" consisting of all cells that immediately neighbor the target cell even by a single point (there are at most 26 of such cells in 3D, and fewer if some of the neighbor cells are not refined as deep as the target cell). The rest of the domain is assigned to the "far zone". All the cells in the far zone that belong to the near zone of the target cell parent but do not belong to the near zone of the target cell itself are labeled as an "interaction zone".

\begin{figure}[t]
\includegraphics[width=\hsize]{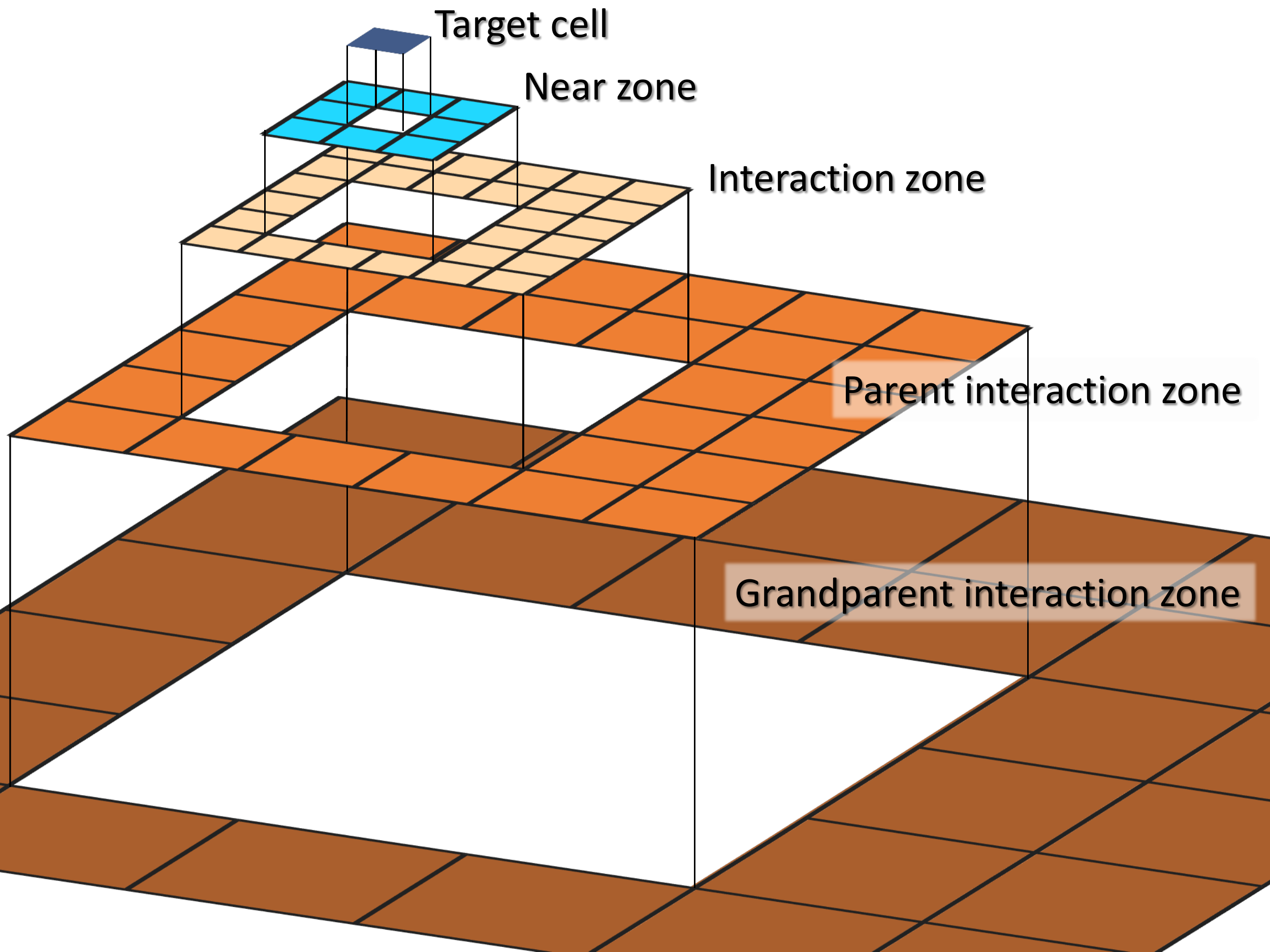}
\caption{\label{fig:fmm} Illustration of the geometry of the FMM. Contributions from the matter in the near zone to the gravitational field at all locations in the target cell are computed directly. Contributions from the interaction zone are computed through their multipole expansions. Contributions from the rest of the far zone are divided into the parent interaction zone plus grandparent interaction zone plus grand-grandparent interaction zone etc.}
\end{figure}

Figure \ref{fig:fmm} shows a sketch of such subdivision of space in 2D (the 3D case is completely analogous). The contribution to the gravitational field everywhere inside of the target cell is composed of 3 contributions.
\begin{enumerate}
    \item The field from the near zone and the target cell itself is computed explicitly, either by direct summation or by some other approach such as non-periodic Fast Fourier Transform (FFT).
    \item The contribution from the interaction zone is computed explicitly using the gravitational field parameterization in the source cells of the interaction zone. In the original FMM terminology this step is called "multipole-to-local" or M2L transformation and reduces to simply computing the gravitational field from the multipole expansion from all interaction zone cells. I will call this step "source-to-target transformation" to allow for non-multipole parameterizations.
    \item The contribution from the rest of the far zone is added through the target cell parent, grandparent, grand-grandparent etc, whose interaction zone contributions summed together make the whole of the rest of the far zone. In fact, if these contributions are accumulated as the tree is traversed from level 0 up (I use the convention when the more refined levels are located "higher" in the tree, as the tree grows "up"), when the parent of a given target cell is ready to add its contribution to the gravitational potential of the target cell, the parent gravitational field already includes all the contributions from the grandparent, grand-grandparent etc all the way down to the root of the tree. I call this last, fourth step of the algorithm "parent-to-child" transformation in contrast to the original FMM terminology of "local-to-local" or L2L translation.
\end{enumerate}
The FMM algorithm is then applied to all leaf cells in the computational domain (or all cells if the gravitational field is required to be known on non-leaf cells too).

The important limitation of the original FMM algorithm is now apparent. In 3D there are up to $6^3-3^3=189$ source cells in the interaction zone, and computing multipole contributions from all of them one by one is going to form a computational bottleneck. It may be more efficient if all these contributions are computed simultaneously, and that is the main feature of the algorithm described below.

\section{Hierarchical Particle-Mesh Method}\label{sec:methods}

Multipole representation of the gravitational field in a target cell from a source cell in the interaction zone is simply a convenient parameterization. A different parameterization would work equally well as long as it provides an equally accurate approximation to the gravitational field. For example, a set of potential values at discrete points on a sphere or cube surrounding the target and source cell to parameterize the internal (for a target cell) or external (for a source cell) gravitational fields have been used extensively in the past \citep[c.f.][]{makino99,km01,ying04,ying06,rogers15,march2017}. Other parameterizations such as Chebyshev polynomials \citep{ChebPoly} or weights of Gaussian quadratures \citep{GaussQuad} have also been used.

\begin{figure}[t]
\includegraphics[width=\hsize]{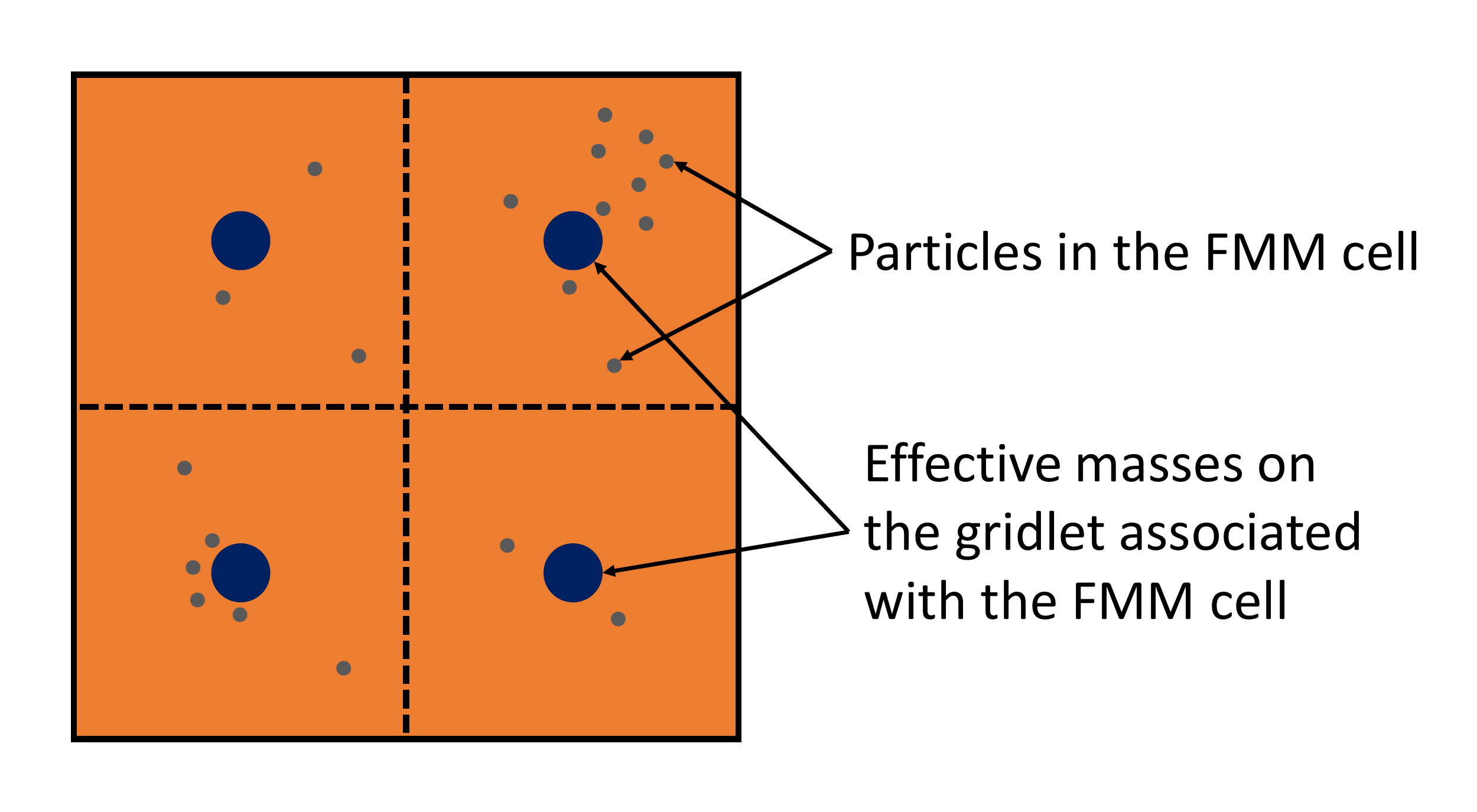}
\caption{\label{fig:ag} Anatomy of a gridlet (shown for a 2D $2\times2$ case). The combined gravitational field of all the particles inside a gridlet is approximated by the gravitational field of effective masses arranged regularly inside the gridlet. The values for the effective masses need to be set carefully for the approximation to be accurate, and this is discussed below in \S \ref{sec:ls}.}
\end{figure}

In the Hierarchical Particle-Mesh (HPM) method the gravitational field is parameterized by effective masses arranged on a small regular grid inside each cell as shown in Figure \ref{fig:ag} - I call such a small grid a "gridlet" in order to distinguish it from the global grid or from a grid covering the interaction zone introduced below. In this terminology, each cell of the FMM tree has an associated unique gridlet and each gridlet covers one and only one cell.

Gridlet-like representation of the multipole expansion and its associated advantage of being able to use FFT is not new and has been used in the past in FMM and non-FMM multipole-based approaches \citep{berman1995,ong03,ying06,liska2014,nitadori14}. The key feature of the HPM method is the use of a \emph{single} FFT for computing the source-to-target transformation for all effective masses (or, equivalently, for all multipole orders) and all cells in the interaction zone at once, and the associated gain in efficiency with FFT padding reduced from the standard factor of 2 to only a factor of $\sfrac{4}{3}$, which is described below. 

Of the previously discussed algorithms, the most similar to HPM is the method described by \citet{liska2014}. HPM would reduce to \citet{liska2014} approach in the special case of (a) uniform distribution of particles on a lattice, and (b) the number of effective masses equal to the number of particles, with each effective mass equal to the corresponding particle mass, and (c) using cell-by-cell FFT for cells in the interaction zone instead of a single full interaction zone FFT. I comment on the relative performance of the two approaches below. A possibility of using FFT for accelerating the source-to-target transformation was also mentioned by \citet{berman1995} but was not described or discussed.

\subsection{Source-to-target ("multipole-to-local") transformation}
\label{sec:s2t}

\begin{figure}[t]
\includegraphics[width=\hsize]{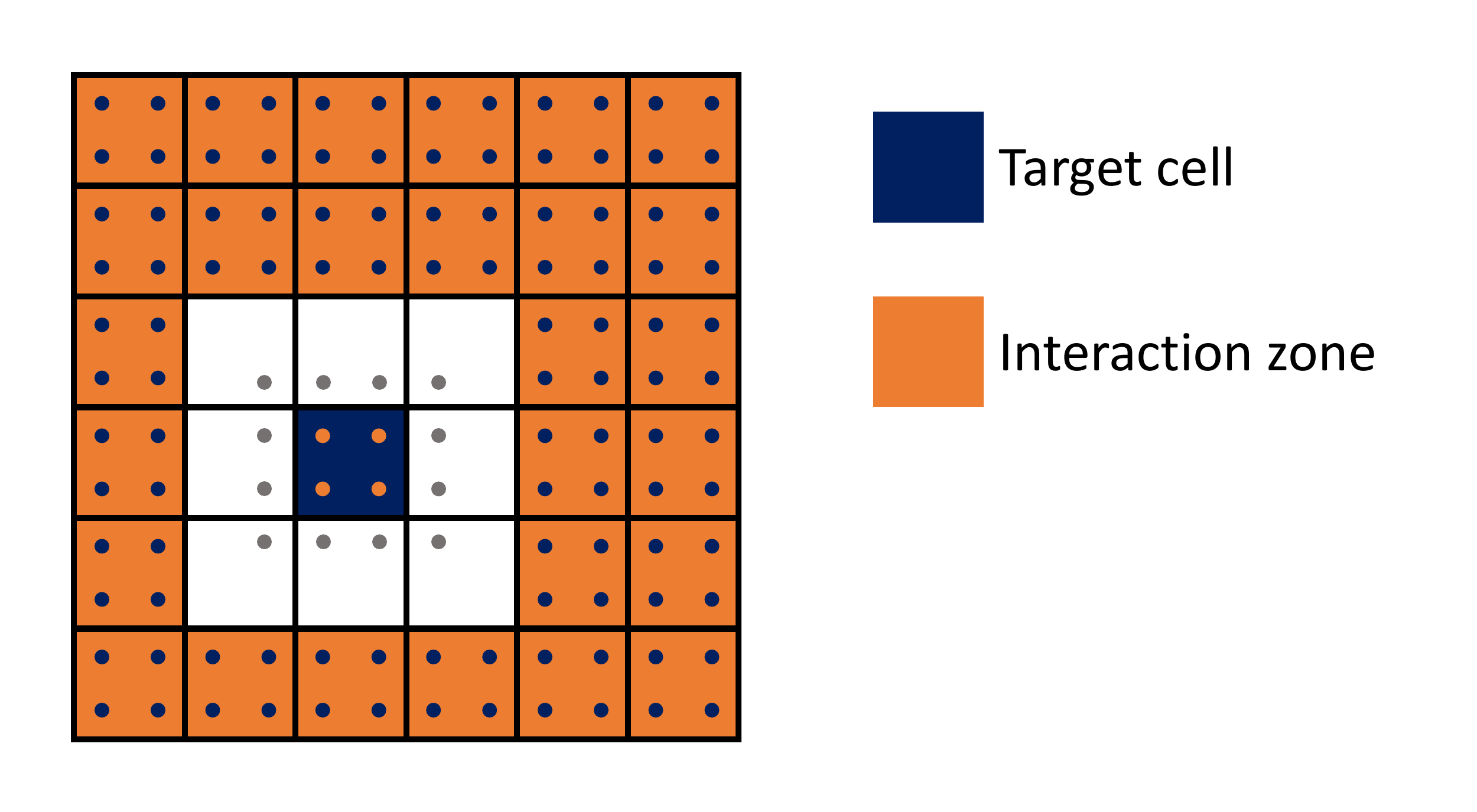}
\caption{\label{fig:iz} Distribution of $2\times2$ gridlets in a 2D case. The effective masses on gridlets are non-zero only in the interaction zone cells. The resulting potential only needs to be computed on the orange gridlet inside the target cell. The grey points belong to an extended gridlet discussed in \S \ref{sec:mods}.}
\end{figure}

Let the number of effective masses on a gridlet along one dimension be $N_g$ ($N_g^D$ effective masses in total for a $D$-dimensional case). Given the distribution of effective masses on the gridlets in the interaction zone, one can compute the source-to-target transformation ("multipole-to-local" transformation in terminology of \citet{gr87,gr97}) for all interaction zone cells at once if one constructs a $(6N_g)^D$ grid of effective masses in the whole interaction zone (dark blue points in Figure \ref{fig:iz} - the size of the interaction zone is always $6^3$ FMM tree cells, as can be seen from Fig.\ \ref{fig:fmm}) and solves for the value of the potential in the target cell gridlet (orange points in Figure \ref{fig:iz}) with a standard particle-mesh (PM) method \citep[c.f.][]{he81} using a fast Fourier Transform (FFT), although with two key caveats. First, the choice of the Green function for the PM solver matters. Using a Green function for a finite difference representation of the Poisson equation would be inaccurate, since the PM solver in that case would retain the low spatial resolution of the gridlet. Instead, one should use the exact Green function for the $1/r$ point mass potential so that the potential values on the gridlet are exact and the multipoles computed from these values are accurate up to the same order as the source multipoles.

Second, and most importantly, the interaction zone does not have periodic boundary conditions, so one needs to use a non-periodic PM solver. The simplest approach to non-periodic FFT is to double the grid size and pad the added volume with zeroes. For example, for the 2D case with $N_g=2$ shown in Fig.\ \ref{fig:iz} one can use a grid of size $N_{\rm tot} = 2\times6\times N_g = 24$ along one dimension. The exact Green function for the Poisson solver can then be constructed as the Fourier transform of the $N_{\rm tot}\times N_{\rm tot}$ array $G_{ij} = \log(|i|_p^2+|j|_p^2)/(4\pi)$, where $|i|_p$ denotes the periodic distance between index $i$ and the origin,
\[
    |i|_p = \left\{\begin{array}{ll}
    i, & \mbox{for $0\leq i\leq N_{\rm tot}/2$}\\
    N_{\rm tot}-i, & \mbox{for $N_{\rm tot}/2< i < N_{\rm tot}$.}
    \end{array}\right.    
\]
The value at $i=0$ and $j=0$ is the "softened" value of the potential; in the tests below I set it to the value at $i=1$ and $j=0$. Obviously, in 3D 
\begin{equation}
    G_{ijk} = \frac{1}{4\pi(|i|_p^2+|j|_p^2+|k|_p^2)^{1/2}}.
    \label{eq:gf}
\end{equation}

An immediate and important optimization can be obtained if one notices that the largest 1D (i.e.\ along-the-axis) distance between the edge of the target cell and any edge of any cell in the interaction zone is only 4, so the Fourier transform only needs to be done in a grid of size $N_{\rm tot} = 8\times N_g$, not $N_{\rm tot} = 12\times N_g$. In that case some of the potential values for gridlets in the interaction zone would be incorrect, but those are discarded anyway (the implied inefficiency of computing and discarding some values is discussed in the next paragraph). Even higher optimization is possible if non-periodic FFT is done on the original $(6\times N_g)^D$ grid without any padding. One such method was described by \citet{james77} \citep[see also][]{mag07}. Unfortunately, the James' method is not suitable for HPM, since it is essentially based on the finite difference representation of the Poisson equation rather than on using the exact Green function (\ref{eq:gf}). A new method for non-periodic FFT with the exact Green function will provide indeed a significant optimization for HPM.

Another potential optimization can be considered if one notices that the potential is only needed to be computed inside the target cell, i.e.\ in a small fraction of the FFT grid. However, this gives only a very modest saving because the forward FFT from the real space to Fourier space needs to be done in the full $(8N_g)^3$ grid (3 passes over the data), the multiplication by the Green function also needs to be done in the full grid (1 pass over data), and in the backward transform the first 1D transform over $k_z$ of the Fourier image $\Phi(k_x,k_y,k_z)$ also needs to be done in full (1 pass over data). Only after that the transform over $k_y$ of the intermediate data $\Phi(k_x,k_y,z)$ can be done for $\oct{1}$ of the full $z$-range, and the last transform of $\Phi(k_x,y,z)$ can be done over $\sfrac{1}{64}$ of the full data pass. Hence, the total operation count for such an optimization is $3+1+1+\oct{1}+\sfrac{1}{64}\approx5.14$ versus the total operation count without such optimization of 7, resulting in saving of about 27\%, probably too little to be worthy of the extra complex bookkeeping.

\subsection{Defining effective masses}
\label{sec:ls}

The key to the accuracy of the HPM method is the correct assignment of the effective masses on gridlets. For example, using a standard Nearest-Grid-Point (NGP) or Cloud-In-Cell (CIC) assignment of particle masses into the gridlet density would result in a highly inaccurate approximation to the combined gravitational field (in a language of multipoles, only the monopole and the dipole are preserved in such assignment). 

Instead, for a given gridlet size $N_g$, one can require that the combined gravitational field of $N_g^D$ effective masses is identical to the gravitational field of $N_g^D$ multipoles. This is directly analogous to the definition of "pseudo-particles" by \citet{makino99}. It is more convenient to use Cartesian multipoles in this case, since spherical multipoles in the original \citet{gr87,gr97} formulation are generally complex numbers, and by using Cartesian multipoles one also avoids complex and computationally expensive operations with spherical harmonics.

Below I use concise and clear notation of \citet{vm10}, although Cartesian multipoles were introduced earlier \citep[c.f.][]{cm1,cm2}. Given a density distribution inside an FMM cell $\rho(\vec{r})$, with boldface notation used to denote spatial vectors, one can define Cartesian multipoles as
\begin{equation}
    Q_\vec{n} = \frac{1}{\vec{n}!} \int \rho(\vec{r}) \vec{r}^\vec{n} \dvec{r},
    \label{eq:cm}
\end{equation}
where the choice of notation is such that for 3D vectors $\vec{n}=(n_x,n_y,n_z)$ and $\vec{r}=(x,y,z)$
\[
    \begin{array}{ll}
    \vec{n}! & \equiv n_x!\,n_y!\,n_z!, \\
    \vec{r}^\vec{n} & \equiv x^{n_x}\,y^{n_y}\,z^{n_z}, \mathrm{and}\\
    \dvec{r} & \equiv dx\,dy\,dz.
    \end{array}
\]
In 2D ad 1D cases only the factors with ($n_x,n_y,x,y$) and ($n_x,x$) respectively would be present. In the following, when the dimension of space needs to be quoted explicitly, I consider the 3D case, although the generalization (or, more exactly, reduction) of the equations to 1D and 2D cases is always trivial.

Effective masses $M_\vec{p}$ in a gridlet with $\vec{p}=(i,j,k)$ and $i,j,k$ spanning the range from 0 to $N_g-1$ can now be defined so that the Cartesian multipoles inside the gridlet up to and including the order of $N_g-1$ are faithfully reproduced, i.e.
\begin{equation}
    Q_\vec{n} = \frac{1}{\vec{n}!} \sum_{\vec{p}=0}^{N_g-1} M_\vec{p} \vec{r}_\vec{p}^\vec{n}.  
    \label{eq:qn}
\end{equation}
For a gridlet of physical size $L$, 
\[
    \vec{r}_\vec{p}^\vec{n} = L^\vec{n} \vec{x}_\vec{p}^\vec{n},
\]
where $L^\vec{n} \equiv L^{n_x+n_y+n_z}$ and $\vec{x}_\vec{p}$ are constants for a given gridlet size $N_g$,
\[
    \vec{x}_\vec{p} = \left(\frac{i+\half}{N_g}-\frac{1}{2},\frac{j+\half}{N_g}-\frac{1}{2},\frac{k+\half}{N_g}-\frac{1}{2}\right)
\]
and hence 
\begin{equation}
    \sum_{\vec{p}=0}^{N_g-1} M_\vec{p} \vec{x}_\vec{p}^\vec{n} = \frac{\vec{n}!}{L^\vec{n}} Q_\vec{n}.
    \label{eq:Anp}
\end{equation}
If we treat the vector-valued indices $\vec{p}$ and $\vec{n}$ as "raveled" sequential indices spanning the range from 0 to $N_g^3-1$, then the left-hand-size of Equation (\ref{eq:Anp}) is just a matrix multiplication with the constant matrix $A_{\vec{n}\vec{p}} \equiv \vec{x}_\vec{p}^\vec{n}$. Matrix $A_{\vec{n}\vec{p}}$ is a Vandermonde matrix, has a non-zero determinant, and has a well defined inverse $B_{\vec{p}\vec{n}} \equiv A_{\vec{n}\vec{p}}^{-1}$, which can be precomputed for the whole simulation. Hence, the first step of the FMM algorithm of computing the effective masses on gridlets reduces to computing the Cartesian multipoles and a matrix multiplication:
\begin{equation}
    M_\vec{p} = \sum_\vec{n} B_{\vec{p}\vec{n}} \frac{\vec{n}!}{L^\vec{n}} Q_\vec{n},
    \label{eq:mp}
\end{equation}
which is only insignificantly more computationally expensive than the original FMM method.

\subsection{Child-to-parent ("multipole-to-multipole") translation}

The second step of the FMM algorithm of computing effective masses on the parent cell gridlet from effective masses on child cell gridlets is also straightforward. From equation (9) of \citet{vm10},
\[
    Q_\vec{n}^P = \sum_\vec{m} \frac{\vec{c}^\vec{m}}{\vec{m}!} Q_\vec{n-m}^C,
\]
where $Q_\vec{n}^P$ are Cartesian multipoles in the parent cell and $Q_\vec{n}^C$ are Cartesian multipoles in a child cell whose center if offset from the center of the parent cell by a vector $\vec{c}$. From Equations (\ref{eq:qn}) and (\ref{eq:mp}) it follows that
\[
    M_\vec{p}^P = \sum_\vec{n} B_{\vec{p}\vec{n}} \frac{\vec{n}!}{L^\vec{n}} \sum_\vec{m} \frac{\vec{c}^\vec{m}}{\vec{m}!} \frac{L^\vec{n-m}}{\vec{(n-m)}!} \sum_{\vec{q}} A_{\vec{(n-m)}\vec{q}} M_\vec{q} 
\]
\[
= \sum_{\vec{q}} T_{\vec{p}\vec{q}}^C M_\vec{q}^C,
\]
and matrix
\[
    T_{\vec{p}\vec{q}}^C = \sum_\vec{n} \sum_\vec{m} B_{\vec{p}\vec{n}} \frac{\vec{n}!}{\vec{m}!\vec{(n-m)}!} \frac{(\vec{a}^C)^\vec{n-m}}{2^\vec{n}} A_{\vec{m}\vec{q}}
\]
describes the transformation from a child effective masses to the parent effective masses for a child identified by vector $\vec{a}^C$, the location of the center of the child cell relative to the parent cell in units where the parent cell goes from -1 to 1, $\vec{a}^{C=1}=(-\half,-\half,-\half)$ for the first child, $\vec{a}^{C=2}=(+\half,-\half,-\half)$ for the second child, etc. Matrices $T_{\vec{p}\vec{q}}^C$ are constant and can be precomputed for the whole simulation. 

\subsection{Parent-to-child ("local-to-local") translation}

The final step of the FMM algorithm, the translation of potential values from the parent gridlet to the child gridlet, now becomes almost trivial. The target-to-source transformation using FFT transforms effective masses $M_\vec{p}$ from the interaction zone cells into the potential values $\Phi_\vec{p}$ on the gridlet of the target cell. Given $\Phi_\vec{p}$, one can create a polynomial approximation for the potential anywhere inside the cell associated with the gridlet by computing the potential multipoles first,
\begin{equation}
    V_\vec{n} = \sum_\vec{p} B_{\vec{p}\vec{n}} \Phi_\vec{p}
    \label{eq:c2pmat}
\end{equation}
and then using them for the polynomial approximation for the potential,
\begin{equation}
    \Phi(\vec{x}) = \sum_{n} V_\vec{n} \vec{x}^\vec{n}.
    \label{eq:phix}
\end{equation}
Given a polynomial approximation for the parent cell $\Phi^P(\vec{x})$, parent contribution to the potential values on the child gridlet are reduced to simple evaluation,
\[
    \Phi_\vec{p}^C = \Phi^P\left(\frac{\vec{a}^C+\vec{x}_\vec{p}}{2}\right).
\]
This step is, therefore, identical to the original FMM method with an additional matrix multiplication in Equation (\ref{eq:c2pmat}).

\subsection{Possible modifications}
\label{sec:mods}

One can also consider several straightforward modifications of the HPM method. For example, since the FFT provides solution inside the entire cube that includes the interaction zone, including the whole of the near zone, one can consider using the potential values not only inside the target cell gridlet (orange points in Fig.\ \ref{fig:iz}, but also outside the target cell (gray points in Fig.\ \ref{fig:iz}). If one defines such an extended gridlet, widened on all sides by just one extra point, the main optimization discussed in \S \ref{sec:s2t} - using $(8N_g)^3$ grid for the Fourier transform rather than naively expected $(12N_g)^3$ - remains valid, because in the FFT algorithm the Nyquist frequency is sampled correctly, i.e.\ for an $n$-long FFT all wavenumbers from 0 to $n/2+1$ are not affected by periodic boundary conditions; for the optimized HPM with $n=8N_g$ there are $4N_g+1$ such wavenumbers, and the gravitational potential on all grey points in Fig.\ \ref{fig:iz} is computed correctly.

One may expect that having more potential values in the target gridlet, and, in particular, including values outside of the target cell, would improve the accuracy of interpolation in equation (\ref{eq:phix}). This is indeed so, however the improvement is only modest, since the error in the potential from the interaction zone is dominated by the finite number of effective masses (equal to the number of multipoles) in the source cells. Tests show that the modest gain in accuracy is compromised by the higher computational cost of a wider gridlet. 

Another possible modification that I explored is to use an alternative method for potential interpolation in the target cells. Given the potential values $\Phi_\vec{p}$ in the target gridlet, one can, for example, use spline interpolation instead of the polynomial interpolation of Equation (\ref{eq:phix}). I tried using B-splines for such interpolation, but again found no improvement. 

In fact, the polynomial interpolation achieves higher accuracy than splines unless the spline order is comparable to the polynomial order. While this may sound counter-intuitive, one should remember that interpolation from Equation (\ref{eq:phix}) is actually the Taylor expansion of the exact gravitational potential around the center of the target cell, and the error term in equation (\ref{eq:phix}) scales as the polynomial of degree $(N_g+1)$. Spline interpolation must have the same order for the error term to be competitive.

One may also wonder whether a different FFT-based approach may be more efficient, namely using FFT to compute source-to-target transformation separately for each cell in the interaction zone with appropriately precomputed Green functions - this is an approach taken by \citet{liska2014}. With padding, such approach is not competitive. Indeed, for the full $(8N_g)^3$ grid the operation count is $(8N_g)^3 \log_2\left((8N_g)^3\right) = 1536 N_g^3 \log_2(8N_g)$, ignoring the prefactor in the FFT operation count of $N\log_2N$, which is irrelevant for the relative comparison. Each cell-by-cell padded FFT would take $(2N_g)^3\log_2\left((2N_g)^3\right)$ operations, and with up to 189 of such cell-by-cell FFTs, the total operation count becomes $4536 N_g^3\log_2(2N_g)$. The ratio of the full grid FFT to 189 cell-to-cell FFTs is then 
\[
    0.34\frac{p+3}{p+1},
\]
where $p=\log_2N_g$. For a monopole approximation ($p=0$) the two approaches are comparable, but for any higher order scheme the full grid FFT wins over cell-by-cell ones.

However, if one uses a non-periodic FFT with the exact Green function without padding, cell-by-cell FFTs become more competitive. In that case the full grid FFT count is $(6N_g)^3 \log_2\left((6N_g)^3\right) = 648 N_g^3(p+1.6)$, while cell-by-cell FFTs add up to $189N_g^3\log_2(N_g^3)=567 N_g^3(p+1)$, always beating the full grid FFT albeit by only about 13\%.

Irrespective of whether the single full grid FFT or cell-by-cell FFTs are used, a non-periodic FFT with the exact Green function and without padding would lead to a substantial speed-up of the HPM method by at least a factor of $1536/648\approx2.4$.

\section{Comparison with the original FMM}

\subsection{Operation and memory scaling}

The first and the last steps of the HPM algorithm differ from their original FMM counterparts by only adding a matrix multiplication. Let $M=N_g^3$ be the total number of multipoles and $K$ is the average number of particles (for a particle code) or grid cells (for a grid code) in each FMM tree cell. In the original FMM computing the multipoles from particles/grid cells (the first step of FMM) and computing the gravitational field from multipoles onto particles/grid cells both require $\approx K\times M$ multiply-add operations if all spherical harmonics can be precomputed and cached - this is possible for a grid code, but is not possible for a particle code, which has higher operation count depending on how spherical harmonics are computed.

In the HPM method an extra contribution of $\approx M^2$ multiply-add operations is added due to matrix multiplications. Since the most likely use case for the FMM algorithm is $K\gg M$, the extra matrix multiplication in its HPM variant does not add any significant overhead. Even for $K\sim M$ the overhead is insignificant since these two steps contribute only a small fraction of the overall computational cost. Both algorithms have the same memory requirements ($M$ multipoles per FMM tree cell), since matrix $B_{\vec{p}\vec{n}}$ in Equations (\ref{eq:mp}) and (\ref{eq:c2pmat}) is constant and can be precomputed once and for all.

The second step of the HPM algorithm (child-to-parent transformation) is identical to its original FMM counterpart and requires $M^2$ operations and no extra memory.

\begin{figure*}[t]
\centerline{\includegraphics[width=\hsize]{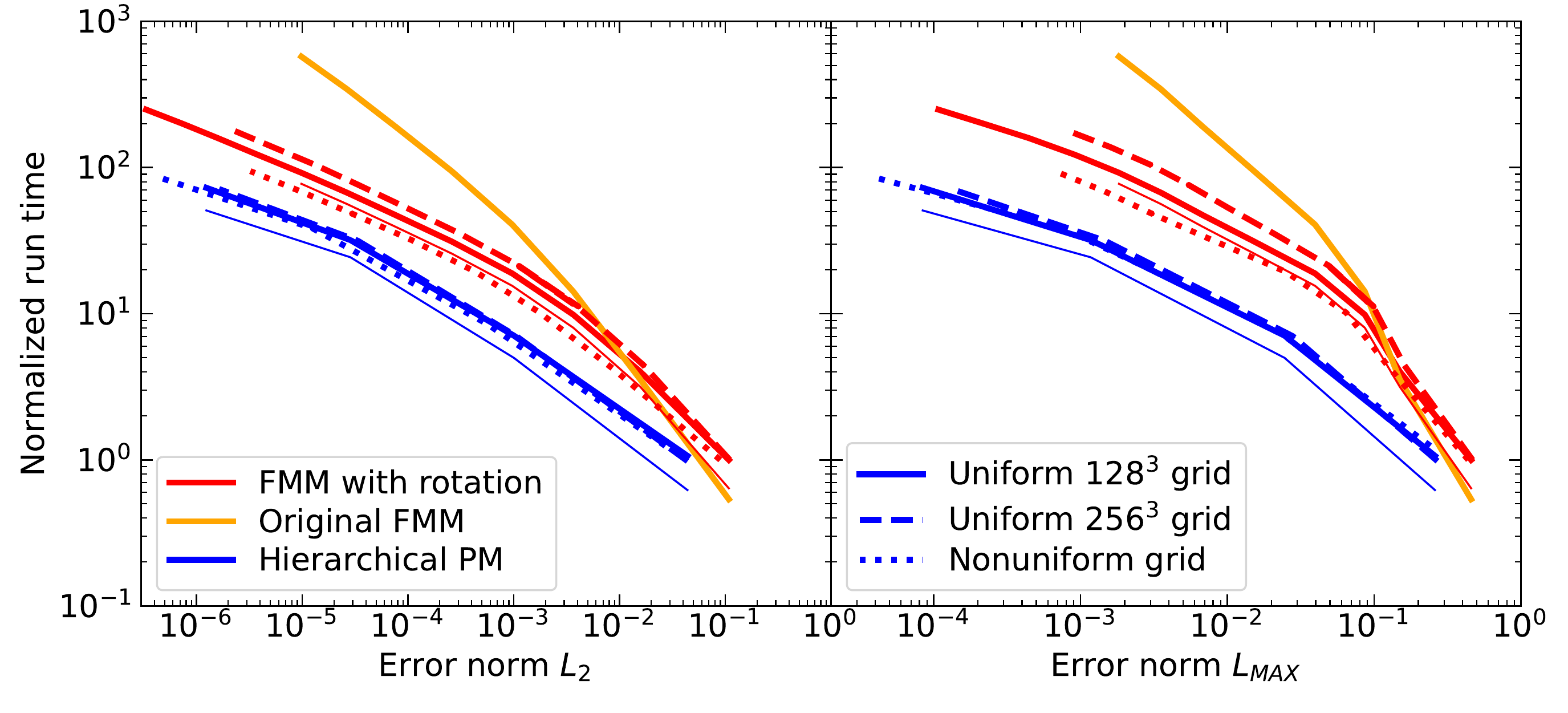}}
\caption{\label{fig:timing} Wall-clock time vs the accuracy of the gravitational force (as measured by $L_2$ and $L_{\rm MAX}$ norms shown in two panels) for the original FMM \citep[][orange]{gr97}, the original FMM with multipole rotation \citep[eq.\ 23 of][red]{cgr99} and the HPM algorithm (blue). Solid, dashed, and dotted lines show three global grid geometries: uniformly refined to 4 levels (solid), uniformly refined to 5 levels (dashed), and uniformly refined to 3 levels with the first octant ($0\leq x,y,z\leq\half$) refined to level 4 and a region ($\oct{1}\leq x,y,z\leq\oct{3}$) refined to level 5. Thin solid red and blue lines match thick solid lines and show timing for the source-to-target transformation step only for the corresponding tests.}   
\end{figure*}

The key difference between the original FMM and HPM is the third, source-to-target transformation. The original FMM algorithm of \citet{gr97} would require at least $5\times189\times M^2\approx950M^2$ multiply-add operations (in a grid code with precomputed spherical harmonics, and even more for a particle code) - the factor of 5 in front comes from the operation count in Equation (5.6) of \citet{gr97}. In order to speed-up this most expensive step of the original FMM algorithm, \citet{gr97} and \citet{cgr99} advocated rotating coordinate frames in which multipoles are computed by aligning the multipole axes with the $z$-axis before transforming multipoles from the source cell to the target cell \citep[eq.\ 23 of][]{cgr99}. Such rotation changes the operation count of the source-to-target transformation to about $2800M^{3/2}$ operations and becomes an optimization for multipoles past the octupole. 

The HPM algorithm as presented only requires about $(8N_g)^3\log_2\left((8N_g)^3\right) = 512 M\log_2(512M)$ operations if the FFT implementation is efficient. If the FFT is done "in place", both the original FMM and HPM require no extra memory.

\subsection{Performance}

For comparison with the original FMM method \citep{gr87,gr97} I use a simple implementation of the HPM algorithm with the source-to-target transformation computed on a $(8\times N_g)^3$ grid (I present comparison for the 3D case as the most common one) using the FFTW solver version 3.3.8 (\url{http://fftw.org}). In order to achieve high precision, the FFT needs to be done in double precision, and all other gridlet values (Cartesian multipoles $Q_\vec{n}$ and $V_\vec{n}$, effective masses $M_\vec{p}$, and the gridlet potentials $\Phi_\vec{p}$) are also kept in double precision. It appears that FFTW works faster in double precision on a 64-bit machine, so going to single precision would not be an optimization.

Since the Poisson equation is linear, it is sufficient to only consider the case of a point source. As the code framework for the comparison I use the "PEx" prototype \citep{gsk18}. PEx implements the oct-tree of small cubic grids ("patches") and solves for gravity using the FMM algorithm on the tree of patches, each patch being a singe "cell" in the FMM tree. Because PEx is the prototype for the hydrodynamic code, each patch in these tests is a grid of $K=8^3$ or $K=16^3$ hydrodynamic cells - the particular value for the patch size is unimportant, the relative performance of the HPM and the original FMM does not depend on the patch size.

The timing of the source-to-parent transformation is not expected to depend significantly on the grid geometry or the level of refinement of the FMM tree. To test for that, I perform 3 different tests: (1) with global grid uniformly refined to 4 levels ($(2^4\times8=128)^3$ underlying uniform grid of hydrodynamic cells), (2) uniformly refined to 5 levels ($256^3$ underlying uniform grid of hydrodynamic cells), and (3) uniformly refined to 3 levels with the first octant ($0\leq x,y,z\leq\half$) refined to level 4 and a region ($\oct{1}\leq x,y,z\leq\oct{3}$) refined to level 5. The accuracy of the numerical solution is measured against the analytically known point source solution for the actual accelerations (the gradient of the potential) with the point source placed at the location (1,1,1); the error is computed as a relative error. The HPM algorithm uses $2^3$, $4^3$, $6^3$, and $8^3$ gridlets, while the original FMM is run with the maximum spherical harmonic number from 1 to 15 with an increment of 2.

In Figure \ref{fig:timing} I show timing for three different algorithms: the original Greengard \& Rokhlin version of FMM \citep{gr97}, the original FMM with multipole rotations, and the HPM flavor of the FMM algorithm as described above versus two different error norms for gravitational accelerations. The HPM method outperforms the original FMM with multipole rotation by a factor of $3-5$ in wall-clock time at a given accuracy or, equivalently, by about an order of magnitude in accuracy for a given wall-clock time.

\section{Conclusions}

The computational performance numbers should be taken with care, as they depend on the specific software implementation. The PEx implementations of all three algorithms are reasonably optimized, with all factors that can be precomputed (spherical harmonics and Wigner matrices in the original FMM at all hydrodynamic cells in a patch, exact Green function and matrices $A_{\vec{n}\vec{p}}$, $B_{\vec{p}\vec{n}}$, and $T_{\vec{p}\vec{q}}$ for the HPM method) and tabulated at the start of the simulation and before the onset of timing. I.e., in all 3 implementations only the operations that explicitly depend on the values of the matter density inside each FMM cell are performed in the timed section of the code, and all other factors, including those that depend only on geometry such as $Y_{lm}(\theta,\phi)$, are precomputed and tabulated. All needed temporary storage is also pre-allocated before the timed section of the code. Never-the-less, no code could not be optimized further.

Another important consideration of any algorithm is parallelization. In both original FMM and HPM all operations are done per FMM tree cell, except in the source-to-target transformation, when multipoles from $189$ source cells are needed for each target cell. Hence, the same parallelization strategy would work for both algorithms. 

The HPM algorithm has another important advantage over the original FMM: since the FFT in the source-to-target transformation is used just to compute the convolution, any Green function in place of Equation (\ref{eq:gf}) would work equally well. Hence, HPM belongs to the class of "kernel independent" variants of the FMM.

\acknowledgments
I am thankful to Volker Springel and Andrey Kravtsov for valuable suggestions and references to related prior work, and to the anonymous referee for the constructive comments that significantly improved the original manuscript.
This manuscript has been authored by Fermi Research Alliance, LLC under Contract No. DE-AC02-07CH11359 with the U.S. Department of Energy, Office of Science, Office of High Energy Physics. 

\bibliographystyle{aasjournal}
\bibliography{main}

\end{document}